\documentclass[a4paper]{article}
\usepackage{graphicx}
\usepackage{amsmath}
\usepackage{amssymb}

\usepackage[margin=1.1in]{geometry}

\usepackage{algorithm}
\usepackage{algorithmic}

\begin{document}
	\title{\bf On the distribution of the sum of dependent standard normally distributed random variables using copulas}
	\author{Walter Schneider\thanks{Berufliche Oberschule Passau, 94032 Passau - Germany, schneider@fos-bos-passau.de}}
	\maketitle
	\pagestyle{headings}

\begin{abstract}
	The distribution function of the sum $Z$ of two standard normally distributed random variables $X$ and $Y$ is computed with the concept of copulas to model the dependency between $X$ and $Y$. By using implicit copulas such as the {\it Gauss-} or {\it t-}copula as well as Archimedean Copulas such as the {\it Clayton-, Gumbel-} or {\it Frank-}copula, a wide variety of different dependencies can be covered.
	For each of these copulas an analytical closed form expression for the corresponding joint probability density function $f_{X,Y}$ is derived. We apply a numerical approximation algorithm in Matlab to evaluate the resulting double integral for the cumulative distribution function $F_Z$. Our results demonstrate, that there are significant differencies amongst the various copulas concerning $F_Z$. This is particularly true for the higher quantiles (e.g. $0.95, 0.99$), where deviations of more than $10\%$ have been noticed.
\end{abstract}
	
\vspace{0.4cm}
	{\bf Keywords: } {\it copula, dependency, distribution function}

\section{Introduction}
Computing the cumulative distribution function (cdf) or probability density function (pdf) of functions of random variables (RV) is a major problem in mathematical statistics. Especially determining the cdf $F_Z$ of the sum $Z = X+Y$ has been extensively studied in the literature.  However the majority of research papers in this area only focus on independent RVs $X$ and $Y$ \cite{AE01, AMH09, DA14}. In case of dependent RVs the most prominent example is given by bivariate normally distributed RVs leading to a normal distribution for the sum $Z$. In this case simple formulas for the mean value $\mu_Z$ and the standard deviation $\sigma_Z$ can be found in basic statistics textbooks, e.g. \cite{PP02}.

Although in many finance or engineering applications, RVs can be assumed to be normally distributed, the dependency amongst the RVs is not necessarily bivariate normal. To overcome the simplifying assumption of bivariate normality, the concept of copulas has been introduced, which allows a flexible modeling of the dependency structure between RVs \cite{NEL06}. The authors in \cite{LPL19} applied copulas for computing the distribution of the product of RVs. Research papers concerning the sum of RVs together with copulas can be found in \cite{DG12, DG13, GH14}. Especially I. Gijbels et al. \cite{GH14} reported results with $X,Y \sim N(0,1)$ and the {\it Gauss-, Gumbel-} and {\it Frank-}copula. Our work is inspired by this paper including the following extensions/differences:

\begin{itemize}
	\item Five copulas are considered: {\it Gauss-, t-, Clayton-, Gumbel-} and {\it Frank-}copula 
	\item The distribution $F_Z$ is computed for different linear correlation coefficients $\rho$ in the range of $0.1 - 0.9$.
	\item Instead of transforming the integration domain, the distribution $F_Z$ is derived in terms of the copula density and the original integration domain.
\end{itemize}

The rest of the paper is organized as follows. In Section 2 we give a brief overview about the theoretical basics of copulas and present the derived bivariate probability density functions $f_{X,Y}$ for all copulas. Then in Section 3 we compute $F_Z$ by numerical integration. The corresponding algorithm is implemented in Matlab, where additionally some useful predefined Matlab copula routines \cite{MAT20} are used. Also, we analyze the differences between the $F_Z$ curves based on all copulas and linear correlation coefficients. Finally we conclude in Section 4.

\section{Theoretical Basics}
As proved in \cite{PP02} the (cumulative) distribution function (df) $F_Z$ of the sum $Z=X+Y$ can be computed by the following proposition. 

\newtheorem{proposition}{Proposition}
\begin{proposition}	
Suppose the random vector $(X,Y)$ has the joint probability density function (jpdf) $f_{X,Y}$, then
\begin{equation}
F_Z(z)=P(X+Y \le z)=\int_{-\infty}^{\infty}\left(\int_{-\infty}^{z-x}f_{X,Y}(x,y)dy\right)dx
\end{equation}
\end{proposition}

In case that the jpdf is not too complicated, a closed-form expression for $F_Z$ can be derived. Otherwise numerical integration algorithms must be used. The major stumbling block however, is the knowledge of a valid jpdf. Only in few cases well known bivariate distributions like the bivariate normal or the bivariate Student-t distribution can be used for modeling the dependency between RVs. Such distributions are restricted to the same type of marginal distributions. Further, they are not suitable for data pairs with some tail dependency. Therefore, a new way of dependency modeling between RVs has been introduced, namely {\it copulas}.

\subsection{The concept of Copulas}
Copulas are helpful in the understanding of dependency at a deeper level, meaning that the dependency is not restricted to the simple case of linear correlation. Copulas combine arbitrary marginal distribution functions to a joint distribution function with a variety of possible dependency models, which are isolated from the marginals. From these dependency models, rank correlations or coefficients of tail dependency can be derived. In the following, the most important basic properties of copulas are summarized. The reader is referred to \cite{NEL06,MFE05} for further detailed information. 

\newtheorem{definition}{Definition}
\begin{definition}
	A $d$-dimensional copula $C$ is a multivariate distribution function on the unit hypercube $[0,1]^d$ with standard uniform marginal distributions, $C:[0,1]^d \rightarrow [0,1]$. The following three properties must hold:
	\begin{enumerate}
		\item $C(u_1,...,u_d)$ is increasing in each component $u_i$.
		\item $C(1,...,1,u_i,1,...,1)=u_i \hspace{0.2cm} \forall i \in \{1,...,d\}, u_i \in [0,1]$.
		\item For all $(a_1,...,a_d),(b_1,...,b_d) \in [0,1]^d$ with $a_i \le b_i$ there is
		\begin{displaymath}
			\sum_{i_1=1}^2 ... \sum_{i_d=1}^2 (-1)^{i_1+...+i_d}C(u_{1i_1},...u_{di_d}) \ge 0
		\end{displaymath}
	    with $u_{j1}=a_j$ and $u_{j2}=b_j \hspace{0.2cm} \forall j \in \{1,...,d\}$
	\end{enumerate}
\end{definition}

In order to work with any type of marginal distributions, the following elementary proposition is necessary.

\begin{proposition}
	Let $G$ be a df and $G^{-1}$ its generalized inverse, i.e. the function $G^{-1}=\inf\{x:G(x) \geq y\}$.
	\begin{enumerate}
		\item {Quantile transformation}. If $U \sim U(0,1)$ has a standard uniform distribution, then 
		\begin{displaymath}
		P(G^{-1}(U) \leq x) = G(x).
		\end{displaymath}
		\item {Probability transformation}. If $Y$ has df $G$, where $G$ is a continuous univariate df, then 
		\begin{displaymath}
		G(Y) \sim U(0,1).
	   \end{displaymath}
	\end{enumerate}
	
\end{proposition}

The construction of multivariate dfs with univariate dfs by using copulas is stated in Sklar's Theorem. A full proof can be found in \cite{SS83}.

\begin{proposition}[Sklar's Theorem]
	Let $F$ be a joint distribution function with marginal distributions $F_1, F_2, ..., F_d$. Then there exists a copula $C$ such that, for all $x_1, x_2, ..., x_d \in \mathbb{R}$, 
	\begin{equation}
		F(x_1,x_2,...,x_d)=C(F_1(x_1),F_2(x_2),...,F_d(x_d))
	\end{equation}
\end{proposition}

Conversely, if $C$ is a copula and $F_1, ..., F_d$ are univariate dfs, then the function $F$ in (2) is a joint df with the marginal distributions $F_1,...,F_d$.

\vspace{0.2cm}
To bridge the gap between $F$ and the bivariate density function $f_{X,Y}$, which is needed in eq.(1), the definition of the copula density $c$ is important.

\begin{definition}
	The copula density $c$ for a continuous copula $C$ is given by partial differentiation
	\begin{equation}
		c(u_1,...,u_d)=\frac{\partial^d}{\partial u_1 ... \partial u_d} C(u_1,...,u_d) \hspace{0.2cm} \forall u \in [0,1]^d.
	\end{equation}
\end{definition}

From Sklar's Theorem the joint density function $f(x_1,...,x_d)$ can now immediately be expressed in terms of the copula density $c$.

\newtheorem{lemma}{Lemma}
\begin{lemma}
	The joint density function $f(x_1,...,x_d)$ is given by
	\begin{equation}
		f(x_1,...,x_d)=c(F_1(x_1),...,F_d(x_d))f_1(x_1)...f_d(x_d)
	\end{equation}
    with $c$ as copula density derived from copula $C$, the marginal distribution functions $F_1,...,F_d$ and the marginal density functions $f_1,...,f_d$.
\end{lemma}

\subsection{Example of Copulas}
In general, copulas can be split into three categories: 1) {\it fundamental} copulas represent a number of important special dependency structures (e.g. independence or perfectly positive (negative) dependency); 2) {\it implicit} copulas are extracted from well-known multivariate distributions with Sklar's Theorem (e.g. Gauss-Copula); 3) {\it explicit} copulas with simple closed-form expressions (e.g. Archimedian copula family). Some of them should be discussed for the bivariate case in more detail.

{\it A1. Gauss-Copula.} For a bivariate normal (Gauss) random vector $(X,Y)$ its copula is a so-called {\it Gauss-Copula} $C_\rho^{Ga}$ with correlation coefficient $\rho$. There is no simple closed form, but $C_\rho^{Ga}$ can be expressed as an integral over the density of $(X,Y)$ as
\begin{align}
	\begin{split}
		C_\rho^{Ga}(u_1,u_2)&=\Phi\left(\Phi^{-1}(u_1),\Phi^{-1}(u_2)\right) \\
		&=\int_{-\infty}^{\Phi^{-1}(u_1)}\int_{-\infty}^{\Phi^{-1}(u_2)} \frac{1}{2\pi \sqrt{1-\rho^2}} \cdot exp \left[\frac{-(x^2-2\rho x y+y^2)}{2(1-\rho^2)} \right] dx dy \\
	\end{split}
\end{align}

with $\Phi^{-1}$ as the inverse of the standard normal df $\Phi$. The Gauss-Copula can be interpreted as a dependency model, which interpolates between perfect positive ($\rho=1$) and negative ($\rho=-1$) dependency.

\vspace*{0.2cm}
{\it A2. t-Copula.} The Student's t-Copula $C_{\rho,\nu}^t$ allows an increased probability of joint extreme events compared to the Gauss-Copula. This copula is given by
\begin{align}
	\begin{split}
		C_{\rho,\nu}^t(u_1,u_2)&=T_{\rho,\nu}\left(T_\nu^{-1}(u_1),T_\nu^{-1}(u_2)\right) \\
		&=\int_{-\infty}^{T_\nu^{-1}(u_1)} \int_{-\infty}^{T_\nu^{-1}(u_2)} \frac{\Gamma(\frac{\nu+2}{2})}{\Gamma(\frac{\nu}{2})\pi\nu\sqrt{1-\rho^2}} \cdot 
		 \left[1+\frac{x^2-2\rho x y + y^2}{\nu(1-\rho^2)} \right]^{-(\nu+2)/2} dxdy  \\
	\end{split}
\end{align}

where $T_{\rho,\nu}$ is the bivariate Student-t df with $\nu$ degrees of freedom and correlation $\rho$; $T_{\nu}^{-1}$ denotes the inverse of the univariate Student-t distribution function $T_{\nu}$.

\vspace*{0.2cm}
{\it A3. Archimedian copula.} Unlike the Gauss- and t-Copula, the Archimedian Copulas have a simple closed form. The {\it Gumbel ($C_\theta^{Gu}$)-}, {\it Clayton ($C_\theta^{Cl}$)-} and {\it Frank ($C_\theta^{Fr}$)-}Copula with the copula parameter $\theta$ are prominent examples for bivariate Archimedian Copulas.
\begin{eqnarray}
		C_\theta^{Gu}(u_1,u_2)&=&exp\left[-\left((-\ln u_1)^\theta+(-\ln u_2)^\theta\right)^{\frac{1}{\theta}}\right], \hspace{0.4cm} 1 \leq \theta < \infty  \\
		C_\theta^{Cl}(u_1,u_2)&=&\left(u_1^{-\theta}+u_2^{-\theta}-1\right)^{-\frac{1}{\theta}}, \hspace{0.4cm} 0 < \theta < \infty \\
		C_\theta^{Fr}(u_1,u_2)&=&-\frac{1}{\theta} \ln\left[1+\frac{(exp(-\theta u_1)-1) \cdot (exp(-\theta u_2)-1)}{exp(-\theta)-1}\right], \hspace{0.4cm}  \theta \in \mathbb{R}
\end{eqnarray}

\begin{figure}[h]
	\includegraphics[height=7.5cm, width=17cm]{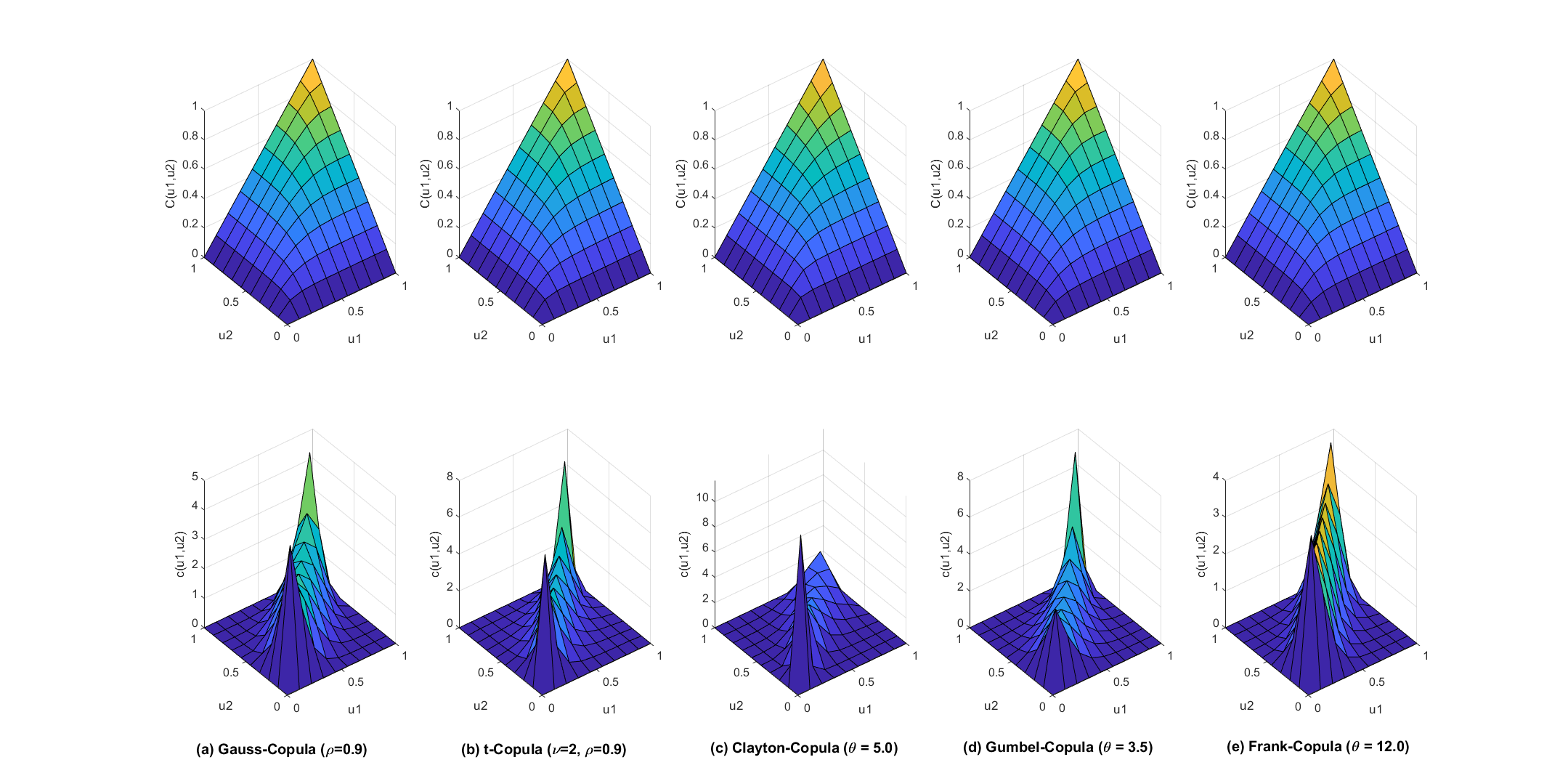}  
	\caption{\small Copula distributions $C$ and densities $c$ for the copulas of Table 1 with $\rho=0.9$ and corresponding $\theta$-parameters}
	\label{fig:copulas}       
\end{figure}

The copula densities $c$ for the introduced copulas are summarized in Table~\ref{tb:copulaDensities} by applying eq.(3) and visualized in Figure~\ref{fig:copulas} together with their distribution functions $C$.

\begin{table}[h]
	\tabcolsep4mm
	\renewcommand{\arraystretch}{2.5}
	\begin{tabular}{|c|c|} \hline
		{\bf Copula $C$}  & {\bf Copula density $c$}  \\  \hline \hline		 
		$C_\rho^{Ga}$ & $ \displaystyle \frac{1}{\sqrt{1-\rho^2}} exp\left[-\frac{\rho^2(x_1^2+x_2^2)-2\rho x_1x_2}{2(1-\rho^2)}\right]$ \hspace{0.5cm} $x_1:=\Phi^{-1}(u_1), x_2:=\Phi^{-1}(u_2)$  \\ \hline
		$C_{\rho,\nu}^t$ & $ \displaystyle \frac{t\left(T_\nu^{-1}(u_1), T_\nu^{-1}(u_2); \nu, \rho\right)}{t_\nu\left(T_\nu^{-1}(u_1)\right)t_\nu\left(T_\nu^{-1}(u_2)\right)}$ \\
		&  $t$: bivariate Student-t density function, $t_\nu$: univariate Student-t density function  \\ \hline
		$C_\theta^{Cl}$ & $\displaystyle (1+\theta)(u_1^{-\theta}+u_2^{-\theta}-1)^{-\frac{1+2\theta}{\theta}}(u_1u_2)^{-(\theta+1)}$ \\ \hline
		$C_\theta^{Gu}$ & $\displaystyle

			\frac{e^{-\sigma_1 } \,{{\left(-\ln\left(u_1 \right)\right)}}^{\theta } \,{{\left(-\ln\left(u_2 \right)\right)}}^{\theta } \,\sigma_1 \,{\left(\theta +\sigma_1 -1\right)}}{u_1 \,u_2 \,\ln\left(u_1 \right)\,\ln\left(u_2 \right)\,{{\left({{\left(-\ln\left(u_1 \right)\right)}}^{\theta } +{{\left(-\ln\left(u_2 \right)\right)}}^{\theta } \right)}}^2 }$ \\			
			& with $\sigma_1={{\left({{\left(-\ln\left(u_1 \right)\right)}}^{\theta} +{{\left(-\ln\left(u_2 \right)\right)}}^{\theta } \right)}}^{1/\theta }$  \\ \hline
			
		$C_\theta^{Fr}$ & $\displaystyle

	\frac{\theta \,{e}^{-\theta \,u_1 } \,{e}^{-\theta \,u_2 } \,{\left({e}^{-\theta \,u_1 } -1\right)}\,{\left({e}^{-\theta \,u_2 } -1\right)}}{{{\left({e}^{-\theta } -1\right)}}^2 \,{\sigma_1 }^2 }-\frac{\theta \,{e}^{-\theta \,u_1 } \,{e}^{-\theta\,u_2 } }{{\left({e}^{-\theta } -1\right)}\,\sigma_1 }$ \\

    & with $
\sigma_1 =\frac{{\left({e}^{-\theta \,u_1 } -1\right)}\,{\left({e}^{-\theta \,u_2 } -1\right)}}{{e}^{-\theta } -1}+1$ \\ \hline
		
	\end{tabular}
	\centering
	\caption{\small Copula density functions for Gauss-, t-, Clayton-, Gumbel- and Frank-Copula}
	\label{tb:copulaDensities}
\end{table} 

The correlation coefficients $\rho$ (Spearman's rho) and  $\tau$ (Kendall's tau) are the most common dependency measures between two RVs and can be expressed in terms of the copula \cite{CZ19}:

\begin{displaymath}
 \tau=4\int_{[0,1]^2}C(u_1,u_2)dC(u_1,u_2)-1,  \hspace{1cm} \rho=12\int_{[0,1]^2}u_1u_2dC(u_1,u_2)-3
\end{displaymath}

From Kendall's $\tau$ the copula parameters $\theta$ for the Archimedean Copulas are given by the following relationships \cite{MFE05}:

\begin{itemize}
\item $C_\theta^{Cl}: \tau=\frac{\theta}{\theta+2}$
\item $C_\theta^{Gu}: \tau=1-\frac{1}{\theta}$
\item $C_\theta^{Fr}: \tau=1-4\theta^{-1}(1-D_1(\theta))$ with the Debye function $D_1(\theta)$.
\end{itemize}

\newtheorem{example}{Example}
\begin{example}
Assuming $\rho = 0.9$ then $\tau=\frac{2}{\pi} \arcsin(\rho) \approx 0.71$ \cite{CZ19}. Then we get the $\theta$-parameters: $\theta(C_\theta^{Cl}) \approx 5.0$, $\theta(C_\theta^{Gu}) \approx 3.5$, $\theta(C_\theta^{Fr}) \approx 12.0$ (see Figure~\ref{fig:copulas}).
\end{example}

\subsection{Copula-based bivariate density functions with standard normal margins}
The bivariate density functions for the RVs $X, Y \sim N(0,1)$ with copula-induced dependency is given according to Lemma 1:

\begin{equation}
	f_{X,Y}(x,y)=c\left(\Phi(x),\Phi(y)\right)\cdot \varphi(x)\varphi(y)
\end{equation}
with pdf $\varphi(x)=\frac{1}{\sqrt{2\pi}}e^{-\frac{x^2}{2}}$ and cdf $\Phi(x)=\frac{1}{\sqrt{2\pi}} \int_{-\infty}^x e^{-\frac{x^2}{2}}$ of the standard normal distribution. The copula densities $c$ are listed in Table 1. After some technical calculations, we obtain the following explicit expressions.\footnote{\footnotesize{Using a computer algebra system (CAS) like Maple, Mathematica or the Symbolic toolbox of Matlab is recommended to verfiy these results.}} 

\begin{itemize}
	\item Gauss-Copula:  
	\begin{equation}
		f_{X,Y}(x,y)=	\frac{1}{2\pi \sqrt{1-\rho^2}} \cdot exp \left[\frac{-(x^2-2\rho x y+y^2)}{2(1-\rho^2)} \right]
	\end{equation}
	It should be noted, that $f_{X,Y}(x,y)$ is identical to the density function of the bivariate standard normal distribution because of its standard normal margins $X, Y$.
	\item t-Copula:
	\begin{equation}
		\begin{array}{l}
			f_{X,Y}(x,y)=\frac{\frac{1}{2\pi \cdot \sqrt{1-\varrho^2 }}\cdot {\left\lbrack 1+\frac{1}{\nu }\frac{{T_{\nu }^{-1} \left(\Phi \left(x\right)\right)}^2 -2\cdot T_{\nu }^{-1} \left(\Phi \left(x\right)\right){\cdot T}_{\nu }^{-1} \left(\Phi \left(y\right)\right)\cdot \rho +{T_{\nu }^{-1} \left(\Phi \left(y\right)\right)}^2 }{1-\rho^2 }\right\rbrack }^{-\frac{\left(\nu +2\right)}{2}} }{\frac{\Gamma \left(\frac{\nu +1}{2}\right)}{\Gamma \left(\frac{\nu }{2}\right)}\cdot \sqrt{\pi \nu }\cdot {\left\lbrack 1+\frac{1}{\nu }\cdot {T_{\nu }^{-1} \left(\Phi \left(x\right)\right)}^2 \right\rbrack }^{-\frac{\left(\nu +1\right)}{2}} \cdot \frac{\Gamma \left(\frac{\nu +1}{2}\right)}{\Gamma \left(\frac{\nu }{2}\right)}\cdot \sqrt{\pi \nu }\cdot {\left\lbrack 1+\frac{1}{\nu }\cdot {T_{\nu }^{-1} \left(\Phi \left(y\right)\right)}^2 \right\rbrack }^{-\frac{\left(\nu +1\right)}{2}} } \cdot \frac{1}{2\pi}e^{\frac{-x^2-y^2}{2}}
		\end{array}
	\end{equation}
	
	\item Clayton-Copula:
	\begin{equation}
		\begin{array}{l}
			f_{X,Y}(x,y)=\left(1+\theta \right)\cdot {\left({\Phi \left(x\right)}^{-\theta } +{\Phi \left(y\right)}^{-\theta } -1\right)}^{-\frac{1+2\theta }{\theta }} \cdot {\left(\Phi \left(x\right)\cdot \Phi \left(y\right)\right)}^{-\left(\theta +1\right)} \cdot \frac{1}{2\pi }e^{\frac{-x^2-y^2}{2}} 
		\end{array}
	\end{equation}

	\item Gumbel-Copula:
	\begin{equation}
		\begin{array}{l}
			f_{X,Y}(x,y)=\frac{e^{-\sigma_1}  \left(-\ln (\Phi(x))\right)^\theta  \left(-\ln (\Phi(y))\right)^\theta  \sigma_1  (\theta + \sigma_1 -1)}{\Phi(x) \Phi(y) \ln(\Phi(x)) \ln(\Phi(y)) \cdot \left( (-\ln(\Phi(x)))^\theta + (-\ln(\Phi(y)))^\theta \right)^2} \cdot \frac{1}{2\pi }e^{\frac{-x^2-y^2}{2}}  
		\end{array}
	\end{equation}
	\hspace{4cm} where $\sigma_1=\left((-\ln(\Phi(x)))^\theta + (-\ln(\Phi(y)))^\theta \right) ^{\frac{1}{\theta}}$.
	
	\item Frank-Copula
	\begin{equation}
		\begin{array}{l}
			f_{X,Y}(x,y)=\left(\frac{\theta \,{\mathrm{e}}^{-\theta \,\Phi(x) } \,{\mathrm{e}}^{-\theta \,\Phi(y) } \,{\left({\mathrm{e}}^{-\theta \,\Phi(x) } -1\right)}\,{\left({\mathrm{e}}^{-\theta \,\Phi(y) } -1\right)}}{{{\left({\mathrm{e}}^{-\theta } -1\right)}}^2 \,{\sigma_1 }^2 }-\frac{\theta \,{\mathrm{e}}^{-\theta \,\Phi(x) } \,{\mathrm{e}}^{-\theta \,\Phi(y) } }{{\left({\mathrm{e}}^{-\theta } -1\right)}\,\sigma_1 } \right) \cdot \frac{1}{2\pi }e^{\frac{-x^2-y^2}{2}} 
		\end{array}
	\end{equation}
	\hspace{3cm} where $\sigma_1 =\frac{{\left({\mathrm{e}}^{-\theta \,\Phi(x) } -1\right)}\,{\left({\mathrm{e}}^{-\theta \,\Phi(y) } -1\right)}}{{\mathrm{e}}^{-\theta } -1}+1$.

\end{itemize}

\subsection{Simulation of Copulas}
To get a better understanding of the non-trivial closed formulas eq.(11) - eq.(15), especially with respect to the dependency between the RVs $X$ and $Y$, we recommend to sample from the copulas. This is achieved by the following algorithm, in which the two Matlab functions {\it copularnd} and {\it norminv} are very helpful. The interested reader is referred to \cite{MFE05} for some further detailled information.

\begin{algorithm}
	\caption{Generating samples $(x,y)$ from a pair of RVs $(X,Y)$ with $X, Y \sim N(0,1)$ and bivariate copula $C$}
	\begin{enumerate}
		\item Generate two standard uniform random vectors $U_1, U_2 \sim U(0,1)$ with size $n$ and copula induced dependency  \\
		
		\hspace{2cm} {\it Matlab}:  \texttt{u=copularnd('Gaussian',rho,n)} \\
		
		\item Transform each of the variables $U_1, U_2$ to a standard normal distribution using the Quantile transformation of Proposition 2. \\
		
		\hspace{2cm} {\it Matlab}:  \texttt{X=[norminv(u(:,1),0,1) norminv(u(:,2),0,1)]}
	\end{enumerate}
\end{algorithm}

\begin{figure}[h]
	\includegraphics[height=7cm, width=16.5cm]{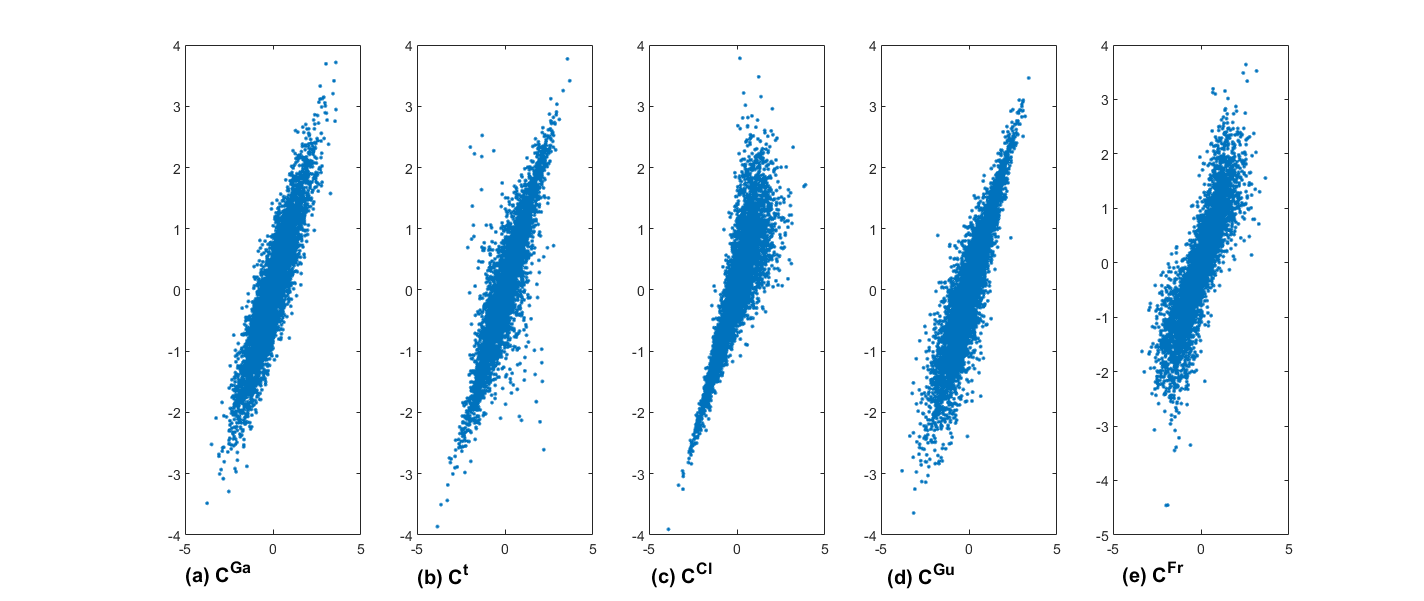}  
	\caption{\small Sampled data pairs for dependent RVs $X, Y \sim N(0,1)$ according to the copulas from Table 1}
	\label{fig:copulaSample}       
\end{figure}

Figure~\ref{fig:copulaSample} shows $5000$ sampled data pairs for the random vector $(X,Y)$. From this illustration we can see, that in contrast to the Gauss-Copula the Clayton-Copula has lower tail dependency, the Gumbel-Copula has upper tail dependency and the t-Copula has both upper and lower tail dependency.

\section{Numerical computation of $F_Z$}

\subsection{Principle of Integration}

\begin{figure}[h]
	\includegraphics[height=8.5cm, width=16cm]{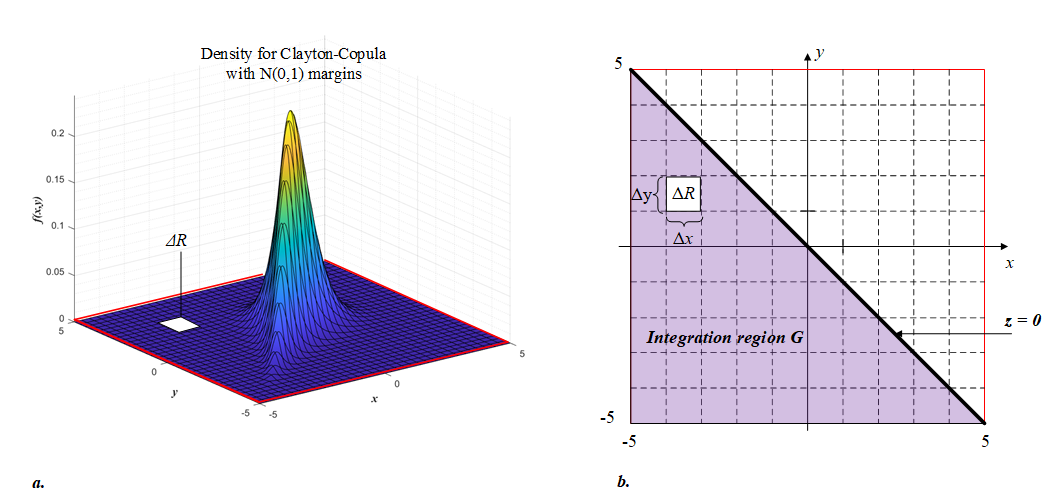}
	\caption{\small a. Density function for the Clayton-Copula with $N(0,1)$ margins, b. construction of the integration region $G$ within the square $[-5;5] \times [-5;5]$ with $x+y\le 0$}
	\label{fig:integ_region}       
\end{figure}

Due to the non-trivial bivariate density functions $f_{X,Y}$, derived in the previous section, an exact analytical expression for the distribution $F_Z$ given by eq. (1) cannot be determined.  So, numerical integration is used instead. As a first approximation, we limit our integration domain to the square $S= [-5;5] \times [-5;5]$ since outside this square $f_{X,Y}$ can be neglected. This is illustrated in Figure~\ref{fig:integ_region}{\it a.} by the example of the Clayton Copula, where the square $S$ is red marked. Numerical calculations have shown that through all discussed copulas $f_{X,Y} < 10^{-6}$ outside $S$. 
So, the distribution function $F_Z$ is approximated by the definite double integral
\begin{equation}
	F_Z(z) \approx \int_{-5}^{5}\left(\int_{-5}^{z-x}f(x,y)dy\right)dx
\end{equation}

with the fixed integration limits of $-5$ and $5$ for the outer integral over $x$.
The inner integral over $y$, however has a variable upper integration limit given by $z-x$. This means that dependent on $z$, we get different integration regions. Figure~\ref{fig:integ_region}{\it b.} shows an example of the integration region $G$ (with violet colour) for $z=0$ within the square $S$.

If we further divide $G$ into a finite set $n \times n$ squares $\Delta R$ with lenghts $\Delta x = \Delta y$ (see Figure~\ref{fig:integ_region}, where the triangles below the straight line $z=0$ can be substituted with squares because of small $\Delta x$) and pick one arbitrary point $P(\xi_i,\eta_i)$ on each $\Delta R$, then we have the following approximation:\footnote{\footnotesize{The exact double integral value is obtained by the sum with $n \rightarrow \infty$}}

\begin{equation}
\int_{-5}^{5}\left(\int_{-5}^{z-x}f(x,y)dy\right)dx \approx  \sum_{j=1}^{n} \sum_{i=1}^{n}f(\xi_i,\eta_i)	 \Delta x \Delta y
\end{equation}

The details of implementing the integral computation are shown in the following Matlab algorithm with $\Delta x = 0.05$. The distribution function $F_Z$ has been computed in the interval $z \in [-5;5]$ with a stepsize of $0.05$.

\begin{algorithm}[h]
	\caption{Approximation of $F_Z$ with numerical computation of the double integral as in eq. (17)}
	\begin{algorithmic}
		\REQUIRE Correlation coefficient $\rho$  
		\STATE $\bullet$ Compute the rank correlation coefficient $\tau$ \\
		\algorithmiccomment{{\it Matlab} \texttt{copulastat:  $\tau$=copulastat('Gaussian',$\rho$)}}
		\STATE $\bullet$ Compute the $\theta$ parameters for Clayton-, Gumbel- and Frank-Copula \\
		\algorithmiccomment{{\it Matlab} \texttt{copulaparam:  $\theta$=copulaparam('family',$\tau$)}}
		\STATE $\Delta x = \Delta y = 0.05$
		\STATE $x, y, z =-5:\Delta x:5$
		\FOR {$k=1:size(z,2)$}
		     \STATE $F_Z$=0
		     \FOR {$i=1:size(x,2)$}
		         \STATE $ylimit=z(k)-x(i)$
	   	         \IF{($ylimit \ge 5$)}
	            	\STATE $yIndex = size(y,2)$
		         \ELSIF{($ylimit \le -5$)}
		            \STATE $yIndex = 1$
		         \ELSE
		         \STATE $yIndex=find(abs(y-ylimit)<0.001)$
		         \ENDIF
		         \FOR{$y=1:yIndex$}
	            	\STATE $f_{X,Y}(i,j)=f_{X,Y}$ using eq.(11) - eq.(15) with $x=x(i)$ and $y = y(j)$
	            	\STATE $F_Z = F_Z + f_{X,Y}(i,j)\cdot \Delta x \Delta y$
		         \ENDFOR
		      \ENDFOR
	          \STATE $F_Z(k)=F_Z$
		\ENDFOR		
	\end{algorithmic}
\end{algorithm}

\subsection{Results and Discussion}

\begin{figure}[h]
       	\includegraphics[height=8cm, width=16cm]{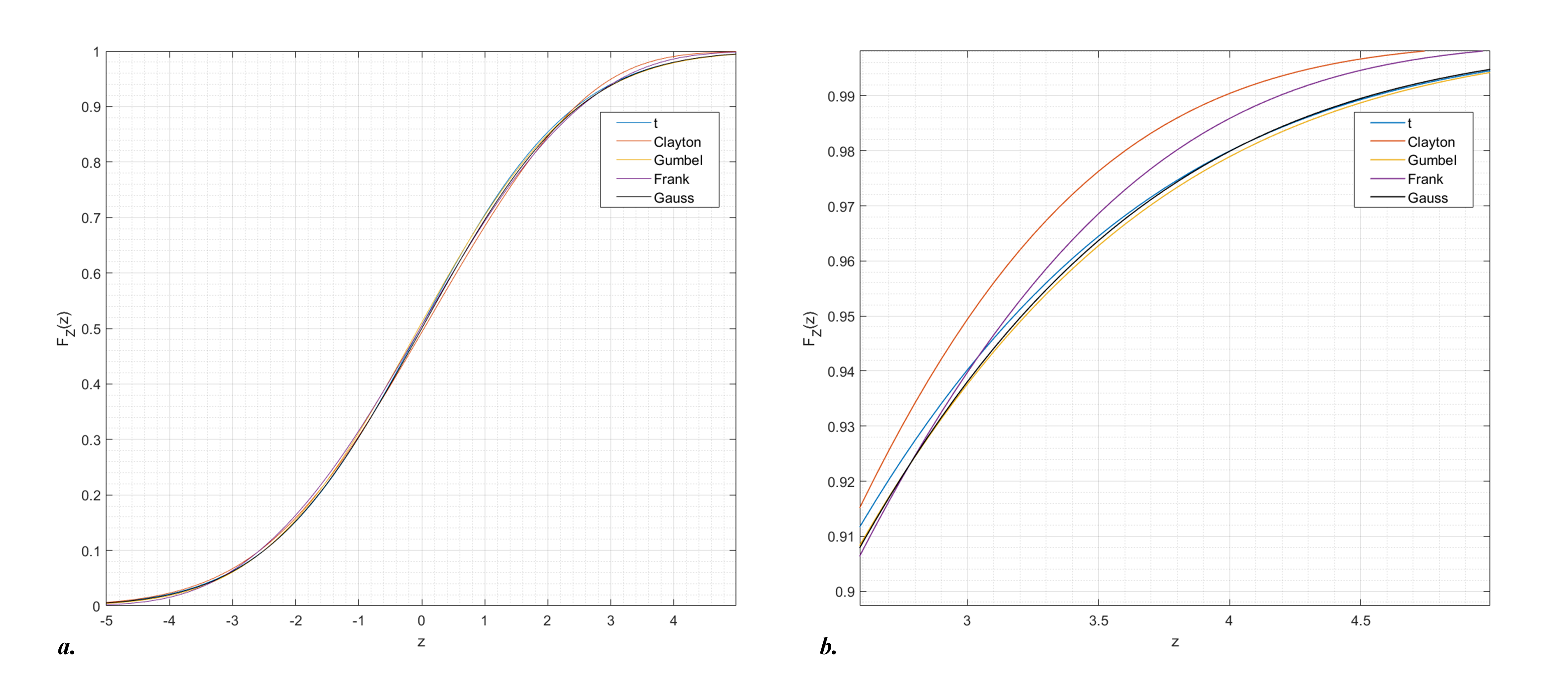}
       	\caption{\small a. Distribution function $F_Z$ of $Z=X+Y$ with $X, Y \sim N(0,1)$, $\rho=0.9$ and different copulas, b. extracted distribution $F_Z$ with quantiles between $0.90$ and $0.99$}
       	\label{fig:CDF_sum}       
\end{figure}

The result of the numerically computed distribution functions with $\rho = 0.9$ and the different copulas is illustrated in Figure~\ref{fig:CDF_sum}. 
Part {\it a.} of the figure shows, that in general the distribution curves are rather similar, but a deeper insight into the higher quantiles (part {\it b.}) reveals, that the Gumbel-Copula is far conservative than the Clayton-Copula: $4.6$ vs. $4.0$ for the $0.99$-quantile, which is a discrepancy of $15\%$ (as seen in Test $\sharp$1 from Table~\ref{tb:copulaResult}). Compared to the Gauss-Copula or equivalently a normal distribution\footnote{\footnotesize{The sum of $X, Y \sim N(0,1)$ with Gauss-Copula induced dependency follows a normal distribution $N(0,\sqrt{2+2\rho})$}} with $\mu=0$ and $\sigma=\sqrt{3.8}$, the $0.99$-quantile for $C^t$ and $G^{Gu}$ are quite similiar (columns 4, 6, 10 of Test $\sharp$1).

\begin{figure}[h]
	\includegraphics[height=8.5cm, width=17cm]{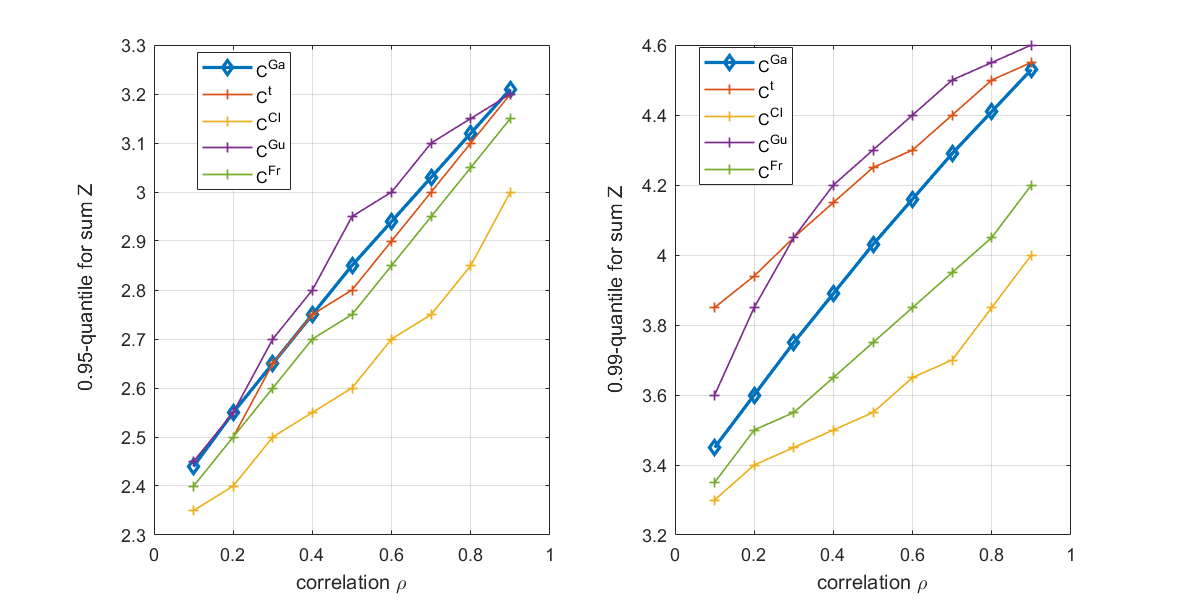}
	\caption{\small $0.95$- and $0.99$-quantiles for the distribution $F_Z$ as function of the correlation $\rho$ and different copulas}
	\label{fig:table_Result}       
\end{figure}

Also we analyze the impact of different correlation coefficients on the distribution curves $F_Z$. Table~\ref{tb:copulaResult} summarizes the corresponding numerical results for the $0.95$- and $0.99$-quantile $z$-values. Visualizing these numerical values in Figure~\ref{fig:table_Result} underlines, that the Clayton-Copula yields the smallest $0.95$-quantiles for any correlation coefficient, however the Gumbel-Copula yields the largest $0.95$-quantiles. This is very similar to the case of the $0.99$-quantiles with the exception that for $\rho < 0.4$ the t-Copula yields the largest $z$-value.

It becomes clear, that disregarding the dependency structure between $X$ and $Y$ can result in a significant error concerning the higher quantiles of the sum's distribution $F_Z$. This is clearly demonstrated in Test $\sharp$5 with $\rho=0.5$ and the $0.99$-quantile: there is a discrepancy of 0.75 between $C^{Cl}$ and $C^{Gu}$ - an error of $21\%$ with respect to $C^{Cl}$. If we rely on a simple normal distribution, then there is discrepancy of 0.48 ($C^{Ga}$ compared to $C^{Cl}$) - still an error of $14\%$. So, assuming a bivariate normal distribution when standard normal margins are given, can lead to serious problems, especially for value-at-risk forecasts.
       
\begin{table}[b]
	\tabcolsep3mm
	\renewcommand{\arraystretch}{1.3}
	\begin{tabular}{|c|c|c|c|c|c|c|c|c|c|c|c|} \hline
		Test$\sharp$  & Corr. $\rho$ & \multicolumn{2}{c|} {$C^{Ga}$} &  \multicolumn{2}{c|} {$C^t$} & \multicolumn{2}{c|} {$C^{Cl}$} & \multicolumn{2}{c|} {$C^{Gu}$} & \multicolumn{2}{c|} {$C^{Fr}$}  \\ 
		&  &  $q95$ & $q99$ & $q95$ & $q99$ & $q95$ & $q99$ & $q95$ & $q99$ & $q95$ & $q99$    \\ \hline \hline
		1 & 0.9 & 3.21 & 4.53 & 3.20  & 4.55  & 3.00  & 4.00  & 3.20  & 4.60  & 3.15  & 4.20   \\ 
		 
		2 & 0.8 & 3.12 & 4.41 & 3.10  & 4.50  & 2.85  & 3.85  & 3.15  & 4.55  & 3.05  & 4.05    \\ 
		
		3 & 0.7 & 3.03 & 4.29 & 3.00  & 4.40  & 2.75  & 3.70  & 3.10  & 4.50  & 2.95  & 3.95    \\
		
		4 & 0.6 & 2.94 & 4.16 & 2.90  & 4.30  & 2.70  & 3.65  & 3.00  & 4.40  & 2.85  & 3.85    \\
		
		5 & 0.5 & 2.85 & 4.03 & 2.80  & 4.25  & 2.60  & 3.55  & 2.95  & 4.30  & 2.75  & 3.75    \\
		
		6 & 0.4 & 2.75 & 3.89 & 2.75  & 4.15  & 2.55  & 3.50  & 2.80  & 4.20  & 2.70  & 3.65    \\
		
		7 & 0.3 & 2.65 & 3.75 & 2.65  & 4.05  & 2.50  & 3.45  & 2.70  & 4.05  & 2.60  & 3.55    \\
		
		8 & 0.2 & 2.55 & 3.60 & 2.50  & 3.95  & 2.40  & 3.40  & 2.55  & 3.85  & 2.50  & 3.50    \\
		
		9 & 0.1 & 2.44 & 3.45 & 2.40  & 3.85  & 2.35  & 3.30  & 2.45  & 3.60  & 2.40  & 3.35    \\ \hline
		
	\end{tabular}
	\centering
	\caption{\small $0.95$- and $0.99$-quantiles ($q95, q99)$ of the sum $Z=X+Y$ with $X, Y \sim N(0,1)$ and dependency given by the linear correlation coefficient $\rho$ and different copulas}
	\label{tb:copulaResult}
\end{table}

\section{Conclusion}
The distribution function $F_Z$ of the sum $Z=X+Y$ with $X, Y \sim N(0,1)$ has been analyzed using different copulas for describing the dependency between $X$ and $Y$. Starting fom basic copula principles, copula-based bivariate density functions $f_{X,Y}$ with standard normal margins have been derived as exact analytical expressions. Based on these functions, $F_Z$ has been computed by numerical integration with Matlab. Our numerical results with five copulas and correlation coefficients in the range of $0.1-0.9$ have shown the importance of dependency modeling. Especially for quantiles of $0.9$ and above significant differences amongst $F_Z$ have been realized.  
In general, the analysis of the dependency of multivariate data sets remains an important task. Future research work will include, that our analysis is extended to the sum of more than two RVs, meaning that high-dimensional copulas must be investigated.

\bibliographystyle{elsarticle-num}
\bibliography{BibJICS}

\begin{thebibliography}{10}
\expandafter\ifx\csname url\endcsname\relax
  \def\url#1{\texttt{#1}}\fi
\expandafter\ifx\csname urlprefix\endcsname\relax\def\urlprefix{URL }\fi
\expandafter\ifx\csname href\endcsname\relax
  \def\href#1#2{#2} \def\path#1{#1}\fi

\bibitem{AE01}
{M. K. Agrawal and S. E. Elmaghraby}, {On computing the distribution function
  of the sum of independent random variables}, in: Computers and Operations
  Research, vol. 28, p. 473 - 483, 2001.

\bibitem{AMH09}
{S. M. Sadooghi-Alvandi, A. R. Nematollahi and R. Habibi}, {On the distribution
  of the sum of independent uniform random variables}, in: Statistical Papers,
  vol. 50, p. 171 - 175, 2009.

\bibitem{DA14}
{C. R. Dance}, {An Inequality for the Sum of Independent Bounded Random
  Variables}, in: Journal of Theoretical Probability, vol. 27, p. 358 - 369,
  2014.

\bibitem{PP02}
{A. Papoulis and S. U. Pillai}, {Probability, Random Variables and Stochastic
  Processes, 4th Edition}, Mc-Graw Hill, 2002.

\bibitem{NEL06}
{R. B. Nelsen}, {An Introduction to Copulas}, Springer Series in Statistics,
  2006.

\bibitem{LPL19}
{Sel Ly, K.-H. Pho, Sal Ly and W.-K. Wong}, {Determining Distribution for the
  Product of Random Variables by Using Copulas}, in: Risks, vol. 7, issue 1,
  2019.

\bibitem{DG12}
{F. Domma and S. Giordano}, {A stress-strength model with dependent variables
  to measure household financial fragility}, in: Statistical Methods
  Application 21 (3), p. 375-389, 2012.

\bibitem{DG13}
{F. Domma and S. Giordano}, {A copula-based approach to account for dependence
  in stress-strength models}, in: Statistical Papers, vol. 54, p. 807-826,
  2013.

\bibitem{GH14}
{I. Gijbels and K. Herrmann}, {On the distribution of the sum of random
  variables with copula-induced dependence}, in: Insurance: Mathematics and
  Economics 59, p. 27-44, 2014.

\bibitem{MAT20}
\href{http://www.mathworks.de}{[link]}.
\newline\urlprefix\url{http://www.mathworks.de}

\bibitem{MFE05}
{A. J. McNeil, R. Frey and P. Embrechts}, {Quantitative Risk Management},
  Princeton University Press, 2005.

\bibitem{SS83}
{B. Schweizer and A. Sklar}, {Probabilistic Metric Spaces}, New York:
  North-Holland/Elsevier, 1983.

\bibitem{CZ19}
{C. Czado}, {Analyzing Dependent Data with Vine Copulas}, Lecture Notes in
  Statistics 222, Springer, 2019.

\end{thebibliography}







\end{document}